\documentclass[aps,pra,twocolumn]{revtex4-1}
\usepackage[latin9]{inputenc}
\usepackage{amsmath}
\usepackage{amssymb}
\usepackage{esint}

\makeatletter
\@ifundefined{textcolor}{}
{%
 \definecolor{BLACK}{gray}{0}
 \definecolor{WHITE}{gray}{1}
 \definecolor{RED}{rgb}{1,0,0}
 \definecolor{GREEN}{rgb}{0,1,0}
 \definecolor{BLUE}{rgb}{0,0,1}
 \definecolor{CYAN}{cmyk}{1,0,0,0}
 \definecolor{MAGENTA}{cmyk}{0,1,0,0}
 \definecolor{YELLOW}{cmyk}{0,0,1,0}
}

\usepackage{epsfig}
\usepackage{bm}
\usepackage{amsfonts}
\usepackage{color}
\usepackage{bigints}
\usepackage{ulem}

\usepackage{amsmath}
\usepackage{latexsym}
\usepackage{mathrsfs}
\usepackage[sans]{dsfont}

\newcommand{\bitem}{\begin{itemize}}
\newcommand{\fitem}{\end{itemize}}
\newcommand{\beq}{\begin{equation}}
\newcommand{\eeq}{\end{equation}}
\newcommand{\beqa}{\begin{eqnarray}}
\newcommand{\eeqa}{\end{eqnarray}}

\newcommand{\dd}{\mbox{d}}

\newcommand {\scat} {\mathrm{sc}}  
\newcommand {\RW} {\mathrm{RW}}

\newcommand {\erf} {\mathrm{erf}}

\newcommand {\FWD} {\mathrm{FWD}}
\newcommand {\BWD} {\mathrm{BWD}}

\newcommand {\bg} {\mathrm{bg}}
\newcommand {\abs} {\mathrm{abs}}
\newcommand {\emi} {\mathrm{em}}

\begin{document}

\title{Collective effects in the radiation pressure force}

\author{R. Bachelard}
\email{bachelard.romain@gmail.com}
\affiliation{Instituto de F\'{\i}sica de S\~{a}o Carlos, Universidade de S\~{a}o Paulo, C.P. 369, 13560-970 S\~{a}o Carlos, SP, Brazil}
\author{N. Piovella}
\affiliation{Dipartimento di Fisica, Universit\`{a} degli Studi di Milano, Via Celoria 16, I-20133 Milano, Italy}
\author{W. Guerin}
\author{R. Kaiser}
\affiliation{Universit\'e C\^ote d'Azur, CNRS, INLN, France}

\date{\today}
\begin{abstract}
We discuss the role of diffuse, Mie and cooperative scattering on the radiation pressure force acting on the center of mass of a cloud of cold atoms.
Even though a mean-field Ansatz (the `timed Dicke state'), previously derived from a cooperative scattering approach, has been shown to agree satisfactorily with experiments, diffuse scattering also describes very well most features of the radiation pressure force on large atomic clouds. We compare in detail an incoherent, random walk model for photons and a diffraction approach to the more complete description based on coherently coupled dipoles. We show that a cooperative scattering approach, although it provides a quite complete description of the scattering process, is not necessary to explain the previous experiments on the radiation pressure force.
\end{abstract}
\maketitle

\section{Introduction}

Light scattering in cold atomic clouds is known to yield a number of interesting features. In the steady state regime, multiple scattering induces an effective long-range, repulsive force, which increases the size of the magneto-optical trap (MOT) and limits its spatial density~\cite{Walker:1990,Townsend:1995,Gattobigio:2010,Camara:2014}. In time-resolved experiments, it leads to `radiation trapping', i.e., a long lifetime of the light inside the sample~\cite{Fioretti:1998,Labeyrie:2003,Labeyrie:2005}. These effects (see Ref.~\cite{Baudouin:2014b} for a review) can be well explained by a diffusive description of light transport, or with an incoherent, random-walk model for photons. In this `diffuse scattering' approach, no information on the phase of the scattered light is required to obtain a satisfactory description of the observed experimental results.

On the contrary, in the end of the 90s, it has been shown that multiple scattering of light by cold atoms can exhibit more subtle features related to the coherence of the scattered light~\cite{Labeyrie:1999,Bidel:2002}. The corresponding experiments were performed using a single probe beam and the connection between the scattered far field and mesoscopic properties like coherent backscattering~\cite{Lagendijk:1996,Rossum:1999,AkkermansMontambaux} has been widely studied in the beginning of the 2000s~\cite{Kaiser:2005,Labeyrie:2008}.

\begin{figure}[!b]
\centering \includegraphics[width=1\linewidth]{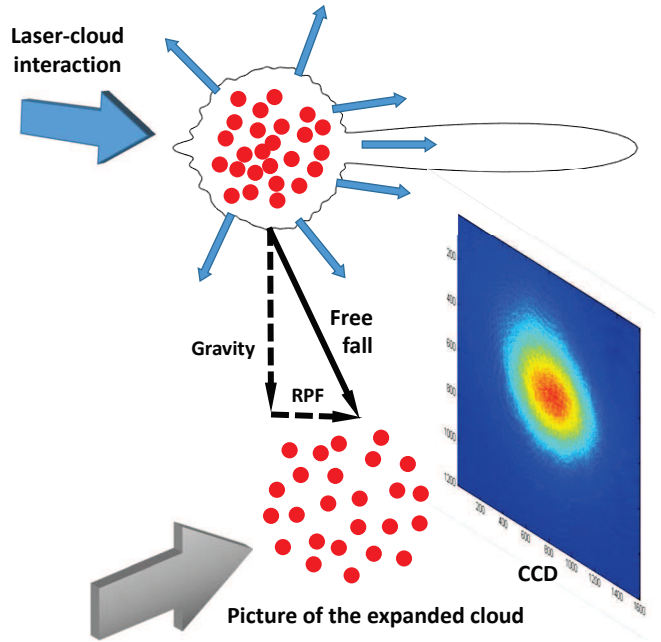}
\caption{\label{fig:scheme} (Color online) Scheme of principle of RPF measurement experiments. After trapping and cooling of the atoms, the produced cloud is released and illuminated by a probe beam for some time. The cloud then realizes a free fall expansion, the motion of each resulting from both gravity and the momentum acquired during the interaction with the probe. At later time, a picture of the expanded cloud is realized, that allows to capture the center of mass displacement due to the light scattering.}
\end{figure}

Starting from an independent approach, cooperative scattering in the low-intensity (or `single-photon') regime, related to Dicke states~\cite{Dicke:1954}, has been investigated in the mid 2000s, first from a theoretical point of view~\cite{Scully:2006}, followed by experiments with presently ongoing efforts in many groups~\cite{Roehlsberger:2010,Keaveney:2012,Meir:2014,Oliveira:2014,Pellegrino:2014,Goban:2015,Guerin:2016,Bromley:2016,Jenkins:2016,Jennewein:2016,Araujo:2016,Roof:2016}. One result, which has been described with a cooperative scattering approach, has been the momentum transfer onto the center of mass of a cloud of cold atoms, measured via the radiation pressure force (RPF)~\cite{Bienaime:2010,Bender:2010,Bux:2010}. A scheme of principle of such experiments is depicted in Fig.~\ref{fig:scheme}.

In the context of further studies on cooperativity, comparisons between coherent and incoherent models of light scattering has been developed~\cite{Chabe:2014,Zhu:2016}. In this article, we apply these models to reconsider the interpretation of previous experimental results~\cite{Bienaime:2010,Bender:2010,Bux:2010} and we investigate to what extend cooperative scattering is merely a convenient way to describe experimental results~\cite{Guerin:JMO}. 

In order to be able to compare numerical to analytical results, we will limit most of the discussion in the present work to the first order correction of the RPF with the optical thickness, i.e., when single scattering dominates.

The paper is organized as follows. In Sec.~\ref{sec:Light}, we discuss the different contributions to the emission diagram of the atomic sample, and the connection between the light scattered and the RPF on the cloud center of mass. In Secs.~\ref{sec:RW}--\ref{sec:TDS}, we describe the contributions on the RPF of, resp., diffuse scattering, diffraction, coherent back-scattering and the prediction based on the driven `timed-Dicke state'~\cite{Scully:2006,Courteille:2010,Bienaime:2011}. Finally, we draw our conclusions in Sec.~\ref{sec:ccl} on the contributions captured by the previous measurements presented in Refs.~\cite{Bienaime:2010,Bender:2010,Bux:2010,Chabe:2014}.

\section{Light radiated from a large atomic cloud, and its relation to radiation pressure force\label{sec:Light}}

\subsection{Contributions to the radiation \label{sec:ScatContrib}}

\begin{figure}[t]
\centering \includegraphics[width=1\linewidth]{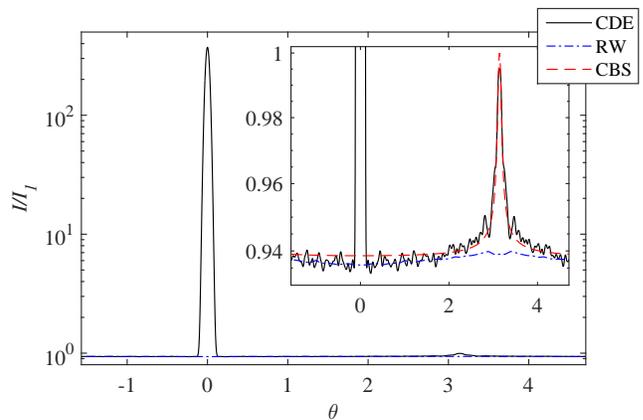}
\caption{\label{fig:radiation} (Color online) Steady-state emission diagram from a Gaussian cloud composed of $N=400$ atoms and with rms radius $kR=25.3$, illuminated by a plane wave with a detuning of $\Delta=\Gamma$. The on-resonance optical thickness is $b_0 = 2N/(kR)^2$ and the actual optical thickness is $b=0.25$. The plain black curve has been obtained from a microscopic coherent scattering model (see Eq.~\eqref{eq:CDE}), the dash-dotted blue one from a random walk of photons simulation (see Sec.~\ref{sec:RW}) and the dashed red curve from a double-scattering computation of the backward radiation (see Sec.~\ref{sec:CBS}). The diagram for the microscopic coherent scattering model has been averaged over $10^4$ realizations.}
\end{figure}

Let us first discuss the light scattered by the cloud since it directly maps to the optical force exerted on the cloud. Fig.~\ref{fig:radiation} depicts a typical radiation pattern for a Gaussian cloud with low optical thickness, illuminated by a plane wave. The radiation pattern has been averaged over many realizations of the atomic positions. Several contributions can be identified: (1) the background radiation, composed of the diffuse scattering by all the atoms (interferences have been averaged out by the configuration averaging), (2) a coherent forward lobe, whose finite angular size comes from the diffraction by the cloud, and (3) the coherent backscattering cone~\cite{Labeyrie:1999,Bidel:2002}, which is a signature of the coherence of light during multiple scattering.

The first contribution is an incoherent one in the sense that the phase of the wave is random from one realization to the other. Consequently the average of the electric field $E$ cancels, yet the average of the intensity $I\propto|E|^2$ does not. This component of the radiation is well described by an incoherent model such as a random walk (RW) of photons~\cite{Labeyrie:2003}, in which phaseless waves are scattered by randomly distributed atoms. This contribution is shown as the dash--dotted blue line in Fig.~\ref{fig:radiation}.

The second contribution corresponds to the diffraction of the incident beam by the cloud, since it acts as a dielectric with a complex refractive index. The effect is coherent in the sense that the scattered wave has a well-defined phase. This part of the radiation is captured by a Mie scattering approach based on a dielectric representation of the sample~\cite{Bachelard:2012}, the finite cloud size leading to a finite angular size of the Forward Lobe (FL). Particle correlations are neglected at this stage~\cite{Saunders:1973b,Morice:1995,Ruostekoski:1997,Javanainen:2016,Bons:2016}. This contribution appears in the plain black curve in Fig.~\ref{fig:radiation} obtained from a microscopic coherent scattering model (see Eq.~\eqref{eq:CDE}).

Finally, the coherent backscattering (CBS) contribution (dashed red curve in Fig.\ref{fig:radiation}) corresponds to a multiple-scattering constructive interference, which is robust against configuration averaging specifically in the backward direction. It is a direct consequence of the correlation between the phase that the wave acquires during one double-scattering event and its reciprocal path. With cold atoms, this phenomenon is particularly sensitive to decoherence~\cite{Chaneliere:2004,Labeyrie:2006b}. The fact that it relies on both disorder and phase coherence means that both the RW and the dielectric approaches are not sufficient to describe it, other techniques are necessary.

Assuming that there is no interference between the three contributions, the scattered intensity is written as
\begin{equation}
I_\scat=I_\mathrm{bck}+I_\mathrm{fwd}+I_\mathrm{CBS} \, .\label{eq:Idecomp}
\end{equation}
This assumption is reasonable since Mie scattering describes the physics of forward coherent scattering, CBS is a coherent phenomenon that occurs in the backward direction and the background corresponds to an incoherent radiation in all directions. Based on the above decomposition, we proceed to analyze their contribution to the pressure force that radiation exerts on the cloud's center-of-mass.

\subsection{Connection to the radiation pressure force}

As can be intuitively expected from total momentum conservation arguments, the pattern of the scattered intensity can be directly mapped to the radiation pressure force exerted on the cloud's center-of-mass. In the present work, we focus on the far-field intensity scattered off the cloud, $I_\scat=2c\epsilon_0|E_\scat|^2$, with $c$ the speed of light, $\epsilon_0$ the vacuum permittivity and $E_\scat$ the scattered electric field. In this far-field limit, the relation between the radiation pressure force on the center-of-mass (in the direction of the incident laser) and the scattered light reads~\cite{Bienaime:2014}
\begin{equation}
\frac{F}{F_1}=\frac{1}{4\pi N}\int_0^{2\pi}\dd\phi \int_0^\pi\dd\theta \sin\theta(1-\cos\theta)\frac{I_\scat(\theta,\phi)}{I_1},\label{eq:IF}
\end{equation}
where $(\theta,\phi)$ refer to the angles in spherical coordinates, with the zenith being given by the wavevector of the incident laser. $F_1=\sigma_1 I_0/c$ refers to the single atom force, with $\sigma_1=4\pi/[k^2(1+4\delta^2)]$ the single atom cross section~\cite{note_scalar}, and $I_1=I_0/[k^2r^2(1+4\delta^2)]$ to the single atom scattered intensity, where $I_0$ is the incident intensity of frequency $\omega=ck$, $r\gg\lambda$ is the distance of the detector from the atomic cloud's center-of-mass and $\delta=(\omega-\omega_0)/\Gamma$ is the detuning of the laser frequency from the atomic resonance frequency in units of the atomic linewidth $\Gamma$. Finally, $N$ is the atom number. In the above equation, the $(1-\cos\theta)$ factor corresponds to the projection on the zenith axis of the momentum transferred from the light to the matter during the absorption--emission process.

In atomic clouds, the radiation pressure force is easier to measure than the full radiation pattern, as the momentum distribution of an atomic cloud can be extracted from time-of-flight experiments. Although the present works focuses on the center-of-mass force, obtained from averaging the momentum gained by all the atoms, detailed information can be extracted from the full momentum distribution~\cite{Inouye:1999}.

\section{Contributions to the radiation pressure force\label{sec:RPF}}

Let us now derive the contributions to the RPF of the different scattering processes described in Sec.~\ref{sec:ScatContrib}. We start by introducing a few useful quantities. In this work, we will consider an incident plane wave $E(\mathbf{r})=E_0e^{ikz}$, so that no dipole force is expected from single-atom physics. We focus on spherical clouds with a Gaussian density distribution
\begin{equation}
\rho(r) = \rho_0 e^{-r^2/2R^2},
\end{equation}
as those routinely produced in MOTs or dipole traps of cold atom experiments. The atomic spatial density at the center of the cloud is given by $\rho_0=N/(\sqrt{2\pi}R)^3$. Using the notation $\mathbf{r} = (x,y,z) = (\mathbf{r}_\perp, z)$, the optical thickness at the center of the cloud ($\mathbf{r}_\perp=0$) is defined as
\begin{equation}
b = \sigma_1\int_{-\infty}^\infty \rho(\mathbf{r}_\perp=0,z) dz =\frac{b_0}{1+4\delta^2}.
\end{equation}
where  $b_0=2N/(kR)^2$ is the resonant optical thickness~\cite{note_scalar}.

\subsection{Diffuse scattering\label{sec:RW}}


Diffuse scattering affects the RPF in two ways: The first one is the so-called shadow effect, that describes the progressive attenuation of the light intensity in the cloud due to the diffuse scattering of light. As a consequence, the first layers of atoms met by the light shield the ones farther in the cloud, resulting in an overall reduced cross section and RPF. This can easily be explained using Beer-Lambert law.

The second effect is a consequence of the first one. The shadow effect implies that more light undergoes a diffusive process near the entrance of the cloud, where it has a higher probability to exit the cloud backward than in the forward direction. Thus, despite each atom scatters the light isotropically, the difference of intensity distribution in the optically thick cloud favors a backward emission~\cite{Labeyrie:2004,Guerin:JMO}. This results in an enhanced radiation pressure force per photon scattered, as compared to a purely isotropic emission. Since the anisotropy of the emission diagram relies on a diffusive process, it can be described by a random walk (RW) model.

Let us now derive more quantitative expressions for these two contributions.

\subsubsection{Shadow effect}

\begin{figure}[!t]
\centering \includegraphics{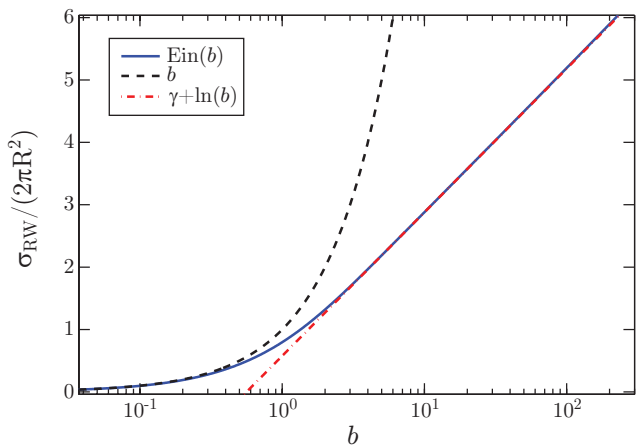}
\caption{\label{fig:sigma_RW} (Color online) Total scattering cross-section of the Gaussian cloud as a function of the optical thickness $b$ [Eq.~(\ref{eq:Srwf})]. For $b\ll 1$, Ein$(b) \simeq b$, while for $b\gg 1$, the cross-section increases logarithmically with $b$, following Ein$(b) \simeq \gamma + \ln(b)$, where $\gamma \approx 0.577$ is the Euler-Mascheroni constant.}
\end{figure}

The first consequence of the shadow effect is a reduction of the total cross section of the cloud. From Beer-Lambert law, the intensity transmission along a line of sight ($\mathbf{r}_\perp$ constant) reads
\begin{eqnarray}
T(\mathbf{r}_\perp) &=& \exp\left(- \sigma_1 \int \rho(\mathbf{r}_\perp,z) dz\right),\nonumber
\\ &=&\exp\left(- b e^{-r_\perp^2/2R^2}\right).\label{eq:Tshadow}
\end{eqnarray}
The cross-section of the cloud corresponds to the part of the light that is scattered, i.e., that is not transmitted, and thus~\cite{Chabe:2014}
\begin{eqnarray}
\sigma_\RW &=&  \int \left[ 1- T(\mathbf{r}_\perp) \right] d^2\mathbf{r}_\perp \nonumber
\\ &=& 2\pi R^2\text{Ein}(b),\label{eq:Srwf}
\end{eqnarray}
with $\text{Ein}(b)$ the entire function
\begin{equation}
\text{Ein}(b)=\int_0^b(1-e^{-t})\frac{dt}{t}=b\left[1+\sum_{n=1}^\infty\frac{(-b)^n}{(n+1)(n+1)!}\right].\label{Ein}
\end{equation}
This function and its asymptotes are plotted in Fig.~\ref{fig:sigma_RW}. The RW cross-section in the small $b$ limit has the following expansion:
\begin{equation}
\sigma_\RW = N \sigma_1 \left( 1-\frac{b}{4}+\frac{b^2}{18}+ \mathcal{O}(b^3) \right) .\label{eq:SrwExp}
\end{equation}
In the limit of vanishing optical thickness $b$, the value expected from single atom physics is recovered, $\sigma_\RW=N\sigma_1$. The deviation from single atom physics corresponds to the shadow effect. For high optical thickness, the cross-section increases only logarithmically, which appears as a saturation of the scattered light~\cite{Pellegrino:2014,Guerin:JMO}.

Consequently, if the light were radiated isotropically (i.e. $I_\scat(\theta,\phi)=I_1\sigma_{RW}/\sigma_1$) then the correction of the shadow effect to the RPF would read
\begin{equation}
\frac{F_\mathrm{Shad}}{F_1}=\frac{\text{Ein}(b)}{b}=1-\frac{b}{4}+\frac{b^2}{18}+ \mathcal{O}(b^3).\label{eq:Fshad}
\end{equation}

\subsubsection{Anisotropy of the emission pattern}

However, as the optical thickness increases, more light escapes the medium in the backward direction than in the forward direction~\cite{Labeyrie:2004}. This comes from the fact that the light intensity is stronger at the entrance of the cloud, where the photons have a higher probability to escape the cloud  backward since the optical thickness in that direction is smaller. Such anisotropy of the emission diagram results in an increase of momentum transferred to the cloud, and so of the radiation pressure force.

Nevertheless this is an higher-order effect in $b$ as compared to the shadow effect, since it requires to account, apart from the attenuating intensity as the light propagates in the cloud, for the probability of the photons to be scattered again after their first scattering event. As shown in Appendix~\ref{Appendix_RW_emission_diagram}, the difference between the forward-scattered ($\theta=0$) and backward-scattered ($\theta=\pi$) intensity, at the lowest order in $b$, reads
\begin{equation}
\frac{I_\FWD-I_\BWD}{I_1} \sim \frac{b^2}{18}.\label{eq:Ianis}
\end{equation}
Although the exact angular pattern is not analytically known, one may approximate this pattern by a sinusoidal function (see Fig.~\ref{fig:anisotropy}) to deduce from Eq.~(\ref{eq:IF}) the following lowest-order correction for the force,
\begin{equation}
\frac{\delta F_\mathrm{anis}}{F_1} \sim \frac{b^2}{108}.
\end{equation}
Thus, this contribution is only second-order in $b$, and much smaller than the second-order contribution of the shadow effect. This angular anisotropy becomes significant at large optical thickness~\cite{Labeyrie:2004,Guerin:JMO}, as most of the light will be scattered backward, but this is beyond the scope of the present work.

\subsubsection{Force of the diffuse scattering contribution}

\begin{figure}[t]
\centering
\includegraphics{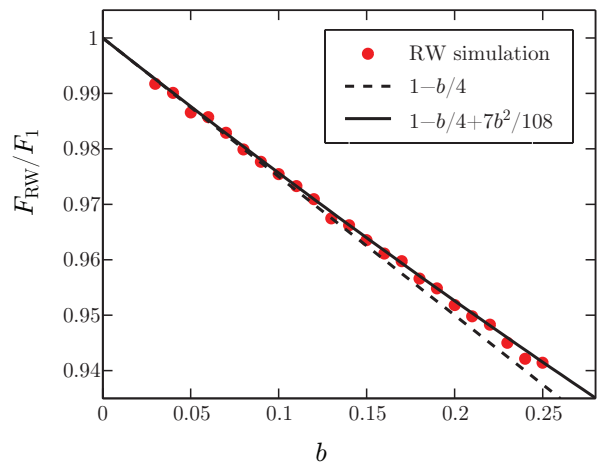}
\caption{(color online). Dependence of the RPF on the optical thickness for the RW model. The correction in $b^2$ to the force can be observed [Eq.~(\ref{eq:F_RW})], yet the $b^2/108$ specific contribution of the anisotropy is beyond the numerical precision of the simulations.}
\label{fig.RWsimu}
\end{figure}

Putting together the contributions of the shadow effect and of the angular anisotropy of diffuse scattering, the
resulting RPF in the RW model reads
\begin{equation}\label{eq:F_RW}
\frac{F_\RW}{F_1} = 1-\frac{b}{4} + \frac{7 b^2}{108}+\mathcal{O}(b^3) .
\end{equation}
We show in Fig.~\ref{fig.RWsimu} the RPF computed with RW numerical simulations, taking into account the Gaussian geometry of the cloud. The decrease of the force due to the shadow effect is well visible, as well as the $b^2$ contribution, up to the computational limitations. 

We thus see that a RW model can produce a reduced RPF on the center of mass of the photons. This reduction takes into account both the attenuation of the incident laser beam in the cloud as well as the modified scattering diagram. As the attenuation inside the cloud dominates the modified emission diagram, the net effect is a reduced RPF for which we derived an analytical expression up to the second order in the optical thickness of the cloud.

\subsection{Diffraction\label{sec:Diff}}

Besides the incoherent diffusion of light discussed in the previous section, the cloud also has a diffracting effect on the incident light due to its finite size. From a macroscopic point of view and as far as diffraction effects are concerned, the atomic cloud can be described as a dielectric medium with an effective complex refractive index, even when the atoms are separated by more than a wavelength.

The real part of the index corresponds to the coherent scattering of the incident beam in the forward direction, with a phase shift, whereas its imaginary part describes the damping of the intensity that accounts for the diffuse scattering of light. In this representation diffuse light is treated as absorption and does not appear any longer in the radiation pattern. 
This representation, usually valid far from resonance, is thus complementary to the RW model, as it discards diffuse light to focus on coherent scattering.

The diffraction by optically-thick three dimensional objects is a challenging problem. Exact solutions exist only for simple geometries and mostly homogeneous systems, following in particular the pioneering work of Gustav Mie~\cite{Hulst:1981}.

On the contrary, in the optically dilute limit, the possibility to resort to a single scattering theory allows studying more complex geometries. Each atom is excited by the incident laser only, and the scattered radiation is the coherent sum of the contributions from each atoms. In this way the complexity of the coupling among the atoms is removed, still preserving interference among scattered radiation by different atoms. As we shall now see, this approach is sufficient to capture the first correction to the RPF due to diffraction, which is valid for small $b_0$ or at large detuning.

In the single scattering limit, the far-field intensity radiated by an ensemble of $N$ atoms at positions $\mathbf{r}_j$, in a direction $\mathbf{\hat k}$ and at a distance $r$, reads
\begin{equation}
I(\mathbf{k})=I_1\sum_{j,m=1}^N e^{-ik(\mathbf{\hat k}-\mathbf{\hat z})\cdot(\mathbf{r}_j-\mathbf{r}_m)},
\end{equation}
The macroscopic effect of diffraction is well captured by neglecting disorder, i.e., treating the cloud as a fluid of density $\rho(\mathbf{r})$ and converting sums $\sum_j$ over the atoms into integrals $\int\mbox{d}\mathbf{r}\rho(\mathbf{r})$ over space. We obtain the following radiation pattern for a Gaussian distribution,
\begin{equation}
\frac{I_\mathrm{fwd}}{I_1}=N^2 e^{-4k^2R^2\sin^2 \theta/2}+\mathcal{O}(b).\label{eq:IFL}
\end{equation}
Thus, the radiation pattern exhibits a forward lobe of coherently scattered light, in an angle $\sim 1/kR$ typical of the diffraction by an object of size $R$. Eq.\eqref{eq:IFL} can then be converted into a force by inserting it into \eqref{eq:IF}, leading to the following contribution to the RPF:
\begin{equation}
\frac{F_\mathrm{fwd}}{F_1}=\frac{N}{8(kR)^4}=\frac{b_0}{16(kR)^2}.\label{eq:FFL}
\end{equation}
The diffraction effect on the RPF is thus small for large clouds since the lobe is very narrow and the light only slightly deviated. Remark that while forward scattering should lead to a reduction of the RPF since the photons exchange less momentum with the matter than if they were scattered isotropically, one rather observes an increase in the force. This is due to the fact that this diffraction effect results in an increase of the scattering cross section (see Ref.\cite{Bienaime:2014} for details), that can be related to the extinction paradox~\cite{Brillouin:1949}.
\begin{figure}[!t]
\centering \includegraphics[width=1\linewidth]{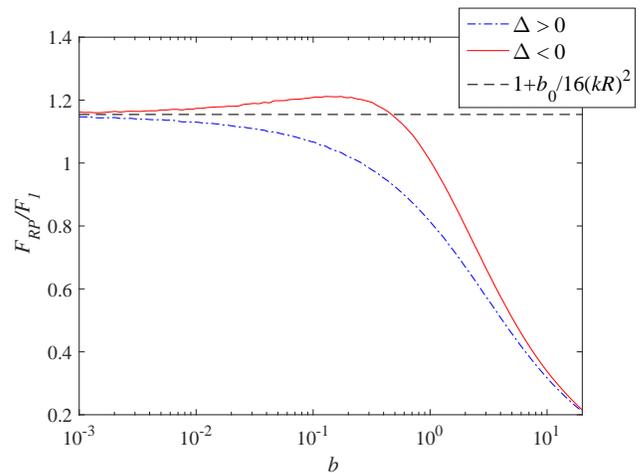}
\caption{\label{fig:diffraction} (Color online) Normalized RPF computed from the coupled-dipole model as a function of the optical thickness $b$ scanned by changing the detuning $\Delta$, for a fixed resonant optical thickness $b_0=22$. The plain red (dash-dotted blue) curve stands for the negative (positive) detuning case. The convergence to the value predicted by diffraction theory (Eq.~(\ref{eq:FFL}), plain black curve) in the $b\to 0$ limit (large detuning) is clearly observed. At intermediate detunings, large phase shifts strongly modify the wave propagation inside the medium, which depends on the sign of detuning $\Delta$~\cite{Roof2015}. The simulations are realized for a cloud with $N=100$, $kR=3$, detuning $\Delta$ from $-75\Gamma$ to $+75\Gamma$, and with an averaging over $10^4$ realizations.}
\end{figure}

Contrary to the other contributions, Eq.\eqref{eq:FFL} does not vanish in the large detuning limit, when $b$ vanishes but not $b_0$ (see Fig.\ref{fig:diffraction}). Since diffraction does not rely on multiple scattering, and since its calculation assumes a coherence between the atomic dipoles just like superradiance~\cite{Araujo:2016,Roof:2016}, it is tempting to interpret this term as a signature of cooperativity and of the role of coherences. However, it simply reflects the lensing effect created by the cloud, as it behaves as a dielectric. Indeed this effect can also be captured using ray optics in dielectrics or Mie scattering (see ref.~\cite{Bachelard:2012} and Appendix~\ref{Modal_expansion}). Finally, the second order contribution to diffraction contains an additional $b$ factor, and its calculation is presented in Appendix~\ref{Modal_expansion}.

\subsection{Coherent Backscattering\label{sec:CBS}}

Let us now analyze the contribution of a process that is considered as a true signature of mesoscopic physics, based on coherent multiple scattering, namely CBS. This coherent effect relies on the constructive interference between a multiple scattering path and its reciprocal path. 

This effect thus requires going beyond single scattering, and also accounting for phase coherence between the scatterers, which is not captured by an incoherent random walk model. We use the Coupled-Dipole Equations (CDE) that describe the all-to-all coupling between the atoms, and the scattering events of any order~\cite{Ruostekoski:1997,Javanainen:1999,Courteille:2010,Svidzinsky:2010,Bienaime:2011}.

We consider $N$ two-level atomic dipoles with a dimensionless amplitude $\beta_j$, at positions $\mathbf{r}_j$, coupled to an incident laser of field $E_0(\mathbf{r})$ and wavenumber $k$ close to the atomic transition (linewidth $\Gamma$), but also coupled to each other through their radiation field:
\begin{equation}
\left(i\Delta-\frac{\Gamma}{2}\right)\beta_j=\frac{id}{2\hbar}E(\mathbf{r}_j)+\frac{\Gamma}{2}\sum_{m\neq j}\frac{\exp(ik|\mathbf{r}_j-\mathbf{r}_m|)}{ik|\mathbf{r}_j-\mathbf{r}_m|}\beta_m.\label{eq:CDE}
\end{equation}
$\Delta$ corresponds to the detuning in frequency between the incident laser and the atomic transition, $d$ the dipole transition element and $\hbar$ the Planck's constant. The field radiated by the cloud at a point $\mathbf{r}$ is
\begin{equation}
E_\scat(\mathbf{r})=-\frac{i\hbar\Gamma}{d}\sum_{j=1}^N \frac{\exp(ik|\mathbf{r}_j-\mathbf{r}|)}{ik|\mathbf{r}_j-\mathbf{r}|}\beta_j.
\end{equation}

By discarding the last term of (\ref{eq:CDE}), this microscopic model allows recovering the single-scattering regime and the diffraction contribution of Eq.~\eqref{eq:FFL}. More generally, the model~\eqref{eq:CDE} contains diffraction for clouds of arbitrary optical thickness and size~\cite{Bachelard:2012}, and describes properly the diffusion of light inside the cloud~\cite{Chabe:2014}. It provides a more general description as compared to the dielectric or the diffusion models, since it also includes effects associated to atomic disorder. Two such examples are the Anderson localization of light~\cite{Skipetrov:2016b} and CBS, which we here discuss. Finally, we note that~\eqref{eq:CDE} describes the light in a scalar approximation, which is a good approximation at low atomic density~\cite{Skipetrov:2014,Bellando:2014}. This limit is relevant to the experiments described in~\cite{Bienaime:2010,Bender:2010,Bux:2010,Chabe:2014}.

The solution of \eqref{eq:CDE} contains scattering at all orders, yet it is in general impossible to calculate it analytically. Indeed only statistical properties of this random matrix can be estimated~\cite{Goetschy:2011b}. A simplifying approach consists in expanding the problem in scattering orders. Whereas the calculation in the previous sections corresponds to single scattering, CBS requires moving at least to second order. We use the decomposition \eqref{eq:Idecomp} for the intensity. Assuming that the background intensity $I_\mathrm{bck}$ is isotropic, its amplitude can be obtained from the cross-section $\sigma_\mathrm{RW}$ as $I_\mathrm{bck}/NI_1=\sigma_\mathrm{RW}/N\sigma_1\approx 1-b/4$ at first order in $b$. For a cloud with Gaussian density, averaging over disorder the scattered intensity up to second order in scattering leads to (see Ref.~\cite{Rouabah:2014} for details):
\begin{eqnarray}
\frac{I_\mathrm{fwd}}{NI_1}&=& N\Big|e^{-2(kR)^2\sin^2(\theta/2)}-\left(1+2i\delta\right)\frac{b}{8}e^{-4(kR)^2\sin^2(\theta/4)}\Big|^2\nonumber
\\ \frac{I_\mathrm{CBS}}{NI_1}&=& \frac{b}{4}\frac{\sqrt{\pi}}{2}\frac{\erf(2kR\cos(\theta/2))}{2kR\cos(\theta/2)},\label{eq:ICBS}
\end{eqnarray}
where the right hand terms in the first line correspond to single and double scattering contributions, whereas the coherent backward contribution of the second line appears only from double scattering~\footnote{Note that in Ref.\cite{Rouabah:2014}, the background contribution from single and double scattering were erroneously assumed to add incoherently (see Eq.(28) of that paper), which resulted in an overestimation of the background intensity. The results on the CBS and forward cone shape and amplitude, as well as the numerical results reported in Fig.5, are however correct.}. The CBS signal is thus enhanced by a factor $2$ as compared to the double scattering background contribution $b/4$ (backscattering corresponds to $\theta=\pi$ and erf$(x)/x \to 2/\sqrt{\pi}$ as $x \to 0$). In the present situation of low optical thickness, it corresponds to an enhancement $b/4$ as compared to the dominant single scattering background. 
In Fig.\ref{fig:radiation}, the contribution from the last line of Eq.\eqref{eq:ICBS} 
 is depicted in dashed red, and presents a good agreement with the CBS signal obtained from the CDE simulations. Of course, this equation does not describe the anisotropy of the emission.

The CBS has an angular width $1/kR$, hence for large clouds the narrowness of this cone makes each photon exchange a momentum of almost $2\hbar k$ with the cloud, thus contributing to the RPF twice as much as a photon scattered in a random direction.
If we focus on the CBS term only (since the contributions of the background and FL were discussed in the previous sections), we obtain the following contribution to the RPF:
\begin{eqnarray}
\frac{F_\mathrm{CBS}}{F_1}&=&\frac{b}{96(kR)^4}\Big[16\sqrt{\pi}(kR)^3\erf(2kR)-12(kR)^2\nonumber
\\ &&+1+(8(kR)^2-1)e^{-4(kR)^2}\Big],
\end{eqnarray}
For a large cloud, the first term dominates and leads to
\begin{equation}
\frac{F_\mathrm{CBS}}{F_1}\approx \frac{\sqrt{\pi}}{6}\frac{b}{kR}.
\end{equation}
The force associated to CBS is positive since it comes from backscattering, yet it is smaller than the negative contribution of the shadow effect by a factor $kR$. For this reason, we were not able to observe it numerically. Another reason may be that while observing CBS already requires averaging over many realizations for optically thick clouds, where the CBS enhancement factor is $2$, much larger numbers of realizations may be required for optically dilute clouds, as the enhancement factor is only $b/2$.

Finally, let us remark that \eqref{eq:ICBS} is valid for two-level atoms. CBS studies on multi-level atoms reported lower enhancement factors~\cite{Labeyrie:1999,Jonckheere:2000}, which would result in a reduced contribution to the RPF.
The CBS contribution to the RPF was thus negligible in the experiments reported in\cite{Bienaime:2010,Bender:2010,Bux:2010,Chabe:2014}.

\subsection{Timed Dicke state\label{sec:TDS}}

The expansion in scattering order discussed previously provides a convenient way to tackle the full many body problem when the optical thickness $b$ is small and few scattering orders are involved. Instead, the approach developed in Ref.~\cite{Courteille:2010} assumed that the atoms acquire the phase given by the laser, with a global amplitude determined by the mean-field solution of Eqs.\eqref{eq:CDE}. This is essentially a mean-field treatment of the power-law coupling present in the scattering problem. This approach is itself inspired by the `Timed-Dicke State'~\cite{Scully:2006} (TDS), an Ansatz that assumes the atoms to acquire the phase of the laser and to have all the same excitation probability. In the limit of large detuning, it successfully captures the hallmark of collective effects, that is, superradiance~\cite{Araujo:2016,Roof:2016}. Let us now discuss which features the TDS captures, among those presented in the above sections.

Following this approach, the RPF for a Gaussian cloud of $N$ atoms reads~\cite{Courteille:2010,Bienaime:2010}
\begin{equation}
\frac{F_\mathrm{TDS}}{F_1}=\frac{4\delta^2+1}{4\delta^2+N^2 s_N^2}Ns_N\left(1-\frac{f_N}{s_N}\right),\label{eq:TDS}
\end{equation}
where we have introduced $s_N$, the angle-averaged structure factor of the cloud emission when illuminated by a plane wave propagating along $\hat{z}$, and a phase function $f_N$ defined as:
\begin{eqnarray}
s_N &=& \langle \langle e^{ik(\mathbf{\hat{z}}-\mathbf{\hat{k}})\cdot\mathbf{r}_j} \rangle_j \rangle_{\mathbf{\hat{k}}},
\\ f_N &=& \langle | \langle e^{ik(\mathbf{\hat{z}}-\mathbf{\hat{k}})\cdot\mathbf{r}_j} \rangle_j |^2 \mathbf{\hat{z}}\cdot\mathbf{\hat{k}} \rangle_{\mathbf{\hat{k}}}.
\end{eqnarray}
and $\mathbf{\hat{k}}$ all the possible scattering directions.
It is then convenient to approximate the cloud by a continuous distribution, yet keeping a disorder contribution to recover the single atom limit~\cite{Bienaime:2010,Bienaime:2011}. Then for large clouds, $kR \gg 1$, these factors read:
\begin{eqnarray}
s_N &\approx& \frac{1}{N}+\frac{1}{4(kR)^2},
\\ f_N&\approx& \frac{1}{4(kR)^2}-\frac{1}{8(kR)^4}.
\end{eqnarray}
These expressions lead to the following formula for the RPF~\cite{note_scalar}
\begin{equation}
\frac{F_\mathrm{TDS}}{F_1}=\frac{4\delta^2+1}{4\delta^2+(1+b_0/8)^2}\left(1+\frac{b_0}{16(kR)^2}\right).
\end{equation}
Let us thus remark that in the large detuning limit, as $b$ vanishes, the TDS predicts a RPF different from the single atom one:
\begin{equation}
\frac{F_\mathrm{TDS}}{F_1}=1+\frac{b_0}{16(kR)^2},
\end{equation}
which precisely corresponds to the diffraction contribution \eqref{eq:FFL}. Since the TDS mimics a single scattering theory modulated by the average change in the dipole population, it makes sense that in the large detuning limit, where the atom population converges to the single-atom-physics one, the results of single scattering theory are recovered.

The other limit of interest is that of small optical thickness $b$. Assuming that $b_0/16\ll 1$, the TDS force can be expanded as
\begin{equation}
\frac{F_\mathrm{TDS}}{F_1}\approx 1-\frac{b}{4},
\end{equation}
which shows that the TDS also contains the first correction of the shadow effect \eqref{eq:Fshad}. This might come as a surprise that this mean-field model takes into account the average attenuation of the laser by the atomic cloud.
Although the TDS neglects the exponential attenuation of the driving field inside the medium and supposes a homogeneous excitation instead, the average value of that excitation corresponds to the first-order shadow effect.

Thus, the TDS is a rather powerful mean-field Ansatz. Initially introduced to describe the superradiant emission of the cloud~\cite{Scully:2006}, it actually captures both the contributions of shadow effect and of diffraction on the RPF at low optical thickness.

\section{Review of the previous experiments \& Conclusions\label{sec:ccl}}

Over the past years, several experiments probed the RPF on atomic clouds. In Ref.~\cite{Courteille:2010}, a timed Dicke state approach was used and the diffraction contribution from the forward lobe Eq.\eqref{eq:FFL} was derived using Eq.\eqref{eq:TDS}. However, the disorder terms was neglected in the structure factors. Consequently, both the single-atom physics and the shadow effect were absent. The contribution of disorder was later included, in Ref.~\cite{Bienaime:2010}, which allowed to recover both single atom physics and the shadow effect, as is presented in Sec.~\ref{sec:TDS}. In Ref.~\cite{Bienaime:2010}, the reduction of the RPF was reported for moderate detunings, where the shadow effect (or its corresponding attenuation described by the TDS Ansatz) is the dominant contribution to the reduced RPF.

The accuracy of the RW approach was further tested in Ref.~\cite{Chabe:2014}, the laser being tuned near the atomic resonance. The RW proved to be fully able to describe the measured RPF in this resonant scattering regime, where diffuse scattering dominates. In that regime, although a model of cooperative scattering can be used [Eq.\eqref{eq:CDE}], addressing the coherences is actually unnecessary.

A measurement of the RPF was also realized far from resonance~\cite{Bender:2010}, and an {\it increase} of the RPF was then observed. The experiment was realized with an ultracold cloud, with high density and small size. This measurement, labeled `cooperative Mie scattering', is here reinterpreted as being the diffraction contribution of the forward lobe \eqref{eq:FFL}, that can be observed only for small dense clouds. Working far from resonance allows making the incoherent scattering contributions negligible.

Finally, in Ref.~\cite{Bux:2010}, another measurement of the RPF was performed out of resonance, and apart from the effects previously described, oscillations in the RPF as the laser detuning was varied were revealed. Up to now, these oscillations remain unexplained by the theories presented in this paper, nor could they be related to Mie resonances phenomena~\cite{Bachelard:2012}. One possibility could be the existence of molecular lines in the excited state of two atoms which affect the scattering cross sections for red detuning and the related RPF.

In conclusion, we have here reviewed the different models used to describe the RPF exerted on the center-of-mass of an atomic cloud, investigating both incoherent and coherent scattering. We have also discussed how recent experiments using coherent scattering models were  actually reporting specific effects that may not always include coherences. In particular, in the papers discussed here, only forward and backward scattering actually require coherence between the dipoles. Their physical mechanisms, namely diffraction and coherent-backscattering, can be understood without a cooperative scattering approach. The collective changes of the radiation pressure force thus do not bear unambiguous signatures of superradiance.



We thank Ph.W. Courteille for stimulating discussions. We acknowledge financial support from the French Agence National pour la Recherche (project LOVE, No. ANR-14-CE26-0032), the Brazilian Funda\c{c}\~ao de Amparo \`a Pesquisa do Estado de S\~ao Paulo and Conselho Nacional de Desenvolvimento Cient\'ifico e Tecnol\'ogico (CNPq, project PVE No. 303426/2014-4).

\appendix

\section{Anisotropy of the emission diagram in the random walk approach}\label{Appendix_RW_emission_diagram}

In this Appendix, we compute the light emission diagram  accounting for its attenuation as it propagates through the cloud, assuming a small optical thickness $b$. 
 The calculation is made in two steps. First, the spatial distribution of the first scattering event is computed from Beer-Lambert law. Then, although the scattering is isotropic, the scattered light has to cross a direction-dependent depth of the medium, depending on (i) the direction of emission and (ii) the scatterer position. The attenuation along this exit path itself feeds higher-order scattering contributions, which we here neglect.

Similarly to the reduction of the cross section by the shadow effect, and since diffraction effects are absent in the RW approach, the transmission up to a point $\mathbf{r}$ of the medium reads
\begin{eqnarray}
T(\mathbf{r}) &=& \exp\left[ -\rho_0 \sigma_1 \int_{-\infty}^{z} e^{-(r_\perp^2+z'^2)/2R^2}dz' \right]
\\ &=& \exp\left\{ -\frac{b}{2} e^{-r_\perp^2/2R^2} \left[ 1+\erf\left(\frac{z}{\sqrt{2}R}\right)\right] \right\}, \nonumber
\end{eqnarray}
with $\erf(z)$ the error function.
The step-length distribution of light scattered in the medium is given by the derivative of the transmission
\begin{eqnarray}
P(\mathbf{r}) & =& -\frac{\partial T}{\partial z}\nonumber
\\ &=& \frac{b}{\sqrt{2\pi}R} e^{-r^2/2R^2}  \exp\Big\{ -\frac{b}{2} e^{-r_\perp^2/2R^2} \nonumber
\\ &&\hspace{1.6cm} \times \left[ 1+\erf(z/\sqrt{2}R)\right] \Big\}.\label{eq:P}
\end{eqnarray}
Note that the integral over space of this step-length distribution allows to recover the cloud cross section
\begin{eqnarray}
\int P(\mathbf{r}) d^3\mathbf{r}&=&-\int (1-T(\mathbf{r}_\perp)) d^2\mathbf{r}_\perp=\sigma_\RW,
\end{eqnarray}
where we have used Eqs.~\eqref{eq:Tshadow} and \eqref{eq:Srwf}. The normalized $P(\mathbf{r})$ gives the position distribution of the first scattering event in the sample.

The attenuation must also be accounted for after the light is scattered. Let us call $T_\mathrm{e}(\mathbf{r},\mathbf{u})$ the transmission from the scattering point $\mathbf{r}$ until the light escapes the cloud in the direction of unit vector $\mathbf{u}$. Taking into account the normalization of $P$, the normalized emission diagram of light is given by
\begin{equation}\label{eq.Iu}
\bar I(\mathbf{u}) = \frac{1}{\sigma_\RW}\int P(\mathbf{r}) T_e(\mathbf{r}, \mathbf{u}) d^3\mathbf{r}.
\end{equation}
$T_\mathrm{e}(\mathbf{r},\mathbf{u})$ is obtained using Beer's law along the escape path of the light
\begin{equation}
T_\mathrm{e}(\mathbf{r}, \mathbf{u}) = \exp\left[ -\frac{b}{\sqrt{2\pi}R} \int_0^\infty e^{-\|\mathbf{r}+t\mathbf{u}\|^2/2R^2} dt  \right],
\end{equation}
at which point it is convenient to use the expansion
\begin{equation}
\|\mathbf{r}+t\mathbf{u}\|^2 = (t+\mathbf{r} \cdot \mathbf{u})^2 + r^2 - (\mathbf{r}\cdot \mathbf{u})^2
\end{equation}
to obtain the expression
\begin{equation}\label{eq.Tru}
T_\mathrm{e}(\mathbf{r}, \mathbf{u}) = \exp\left\{ -\frac{b}{2}  e^{-r^2/2R^2} e^{(\mathbf{r}\cdot \mathbf{u})^2/2R^2} \left[ 1-\erf\left(\frac{\mathbf{r} \cdot \mathbf{u}}{2R^2}\right) \right] \right\}.
\end{equation}

Injecting Eqs.~\eqref{eq:P} and \eqref{eq.Tru} in Eq.~(\ref{eq.Iu}), we obtain the normalized emission diagram under an integral form:
\begin{eqnarray}\label{eq:RadPat}
\bar I(\mathbf{u}) &=&  \frac{bI_1}{\sqrt{2\pi} R\sigma_\RW} \int  e^{-r^2/2R^2}  \nonumber\\
&\times &\exp\left[-\frac{b}{2} \left( f(r_\perp, z) + g(\mathbf{r}, \mathbf{u}) \right)\right] d^3\mathbf{r} \;,
\end{eqnarray}
where we have introduced the following functions:
\begin{align}
f(r_\perp, z) & = e^{-r_\perp^2/2R^2} \left[ 1+\erf(z/\sqrt{2}R)\right]\; ,\\
g(\mathbf{r}, \mathbf{u}) & = e^{-r^2/2R^2} e^{(\mathbf{r}\cdot \mathbf{u})^2/2R^2} \left[ 1-\erf\left(\frac{\mathbf{r}\cdot\mathbf{u}}{\sqrt{2}R}\right) \right] \; .
\end{align}
In order to obtain a tractable expression for the radiation pattern, we expand \eqref{eq:RadPat} to second order in $b$:
\begin{equation}\label{eq:ExpI}
\begin{split}
\bar I(\mathbf{u}) \approx & \frac{b}{\sqrt{2\pi}R\sigma_\RW}\int  d^3\mathbf{r} \; e^{-r^2/2R^2}  \\
&\times \left[1-\frac{b}{2} \left( f+g \right) + \frac{b^2}{8} \left( f^2+ g^2 + 2fg \right)\right] ,
\end{split}
\end{equation}
At this point it becomes clear that the anisotropy of the emission diagram is a second order effect in $b$. Indeed $f$ does not depend on the emission direction $\mathbf{u}$ so its integral over space will not present any anisotropy; $g(\mathbf{r}, \mathbf{u})$ depends only on $\mathbf{r}\cdot\mathbf{u}$ and $r$, and does not present inhomogeneity, so its integral over space is actually isotropic. Consequently only the last $fg$ term in \eqref{eq:ExpI} contributes to the anisotropic emission, as it combines the inhomogeneity of $f$ and the anisotropy of $g$.

The first terms are easily integrated due to the natural separation of the variables:
\begin{eqnarray}
\int e^{-r^2/2R^2} ( f+g ) d^3\mathbf{r}  &=& \left(\sqrt{2\pi}R \right)^3,\\
\int e^{-r^2/2R^2} ( f^2+g^2 ) d^3\mathbf{r} &=& \frac{8}{9}\left(\sqrt{2\pi}R\right)^3.
\end{eqnarray}

\begin{figure}[t]
\centering \includegraphics[width=1\linewidth]{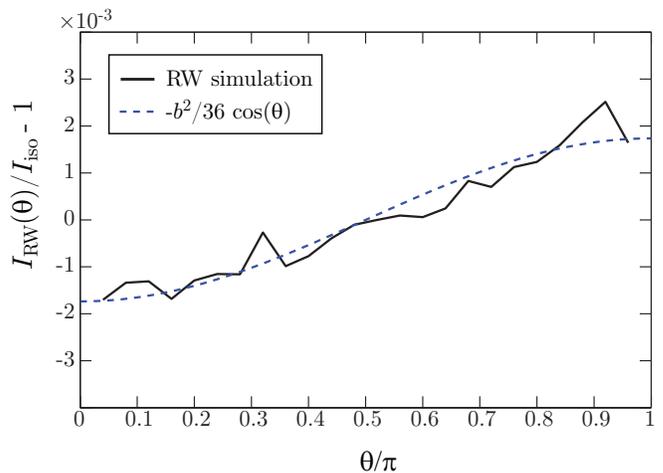}
\caption{\label{fig:anisotropy} (Color online) Emission diagram of a Gaussian cloud of optical thickness $b=0.25$. The black curve corresponds to numerical simulations realized with a RW code, and the dashed blue curve to $I_\RW(\theta) = I_\mathrm{iso} [1-(b^2/36)\cos\theta]$, following Eq.~\eqref{eq:Ian}.}
\end{figure}

The last term in \eqref{eq:ExpI} could not be computed exactly, yet the amplitude of the anisotropy can be obtained by evaluating its difference in the forward ($\mathbf{u}=\hat{z}$) and backward ($\mathbf{u}=-\hat{z}$) directions:
\begin{align}
&\int  e^{-r^2/2R^2} \left[f(r_\perp, z)g(\mathbf{r}, \hat{z})-f(r_\perp, z)g(\mathbf{r}, -\hat{z})\right]  d^3\mathbf{r}\nonumber
\\  & = \frac{2}{9}\left(\sqrt{2\pi}R \right)^3.\label{eq:AnisTerm}
\end{align}
At this point, we approximate the $\theta$-dependence of the emission diagram by a sinusoidal function, i.e., $\bar I(\mathbf{u})\approx I_\mathrm{iso}(b)+\epsilon(b)\mathbf{u}\cdot\hat{z}$. This assumption is supported by RW simulations, see Fig.~\ref{fig:anisotropy}. Then the isotropic contribution of this term is calculated from the average between the backward and forward terms, which leads to $(1/3)(\sqrt{2\pi}R)^3$.

Finally, using Eqs.\eqref{eq:SrwExp} and \eqref{eq:ExpI}--\eqref{eq:AnisTerm}, we can write
\begin{equation}
\bar I(\mathbf{u}) =  1-\frac{b}{4}+\left(\frac{11}{144}-\frac{\mathbf{u}\cdot\hat{z}}{36}\right)b^2+\mathcal{O}(b^3)\label{eq:Ian}
\end{equation}

Inserting the anisotropic term into Eq.~\eqref{eq:IF}, we obtain the contribution of the anisotropy to the RPF:
\begin{equation}
\frac{\delta F_\mathrm{anis}}{F_1} = \frac{b^2}{108}.
\end{equation}

\section{Modal expansion}\label{Modal_expansion}

Light scattering in dielectrics with simple geometries were often investigating using a modal expansion, following the pioneering work of Gustav Mie for homogeneous dielectric spheres~\cite{Hulst:1981}. Atomic clouds can be described as a dielectrics, at least in the regime of linear optics and neglecting particle-particle correlations~\cite{Saunders:1973b,Morice:1995,Ruostekoski:1997,Javanainen:2016,Bons:2016}. It is then possible to use modal expansions to determine the RPF on the center-of-mass of a macroscopic cloud~\cite{Bachelard:2011,Bachelard:2012}.

Here, using the approach of Ref.~\cite{Bachelard:2011}, we show that the modal expansion allows us to recover the first order contribution of the shadow effect and of the forward lobe. One important assumption of this model is that the system is a thin phase object, such that there is no significant phase shift as the light propagates in the system, so that the imaginary part of the interaction kernel $\exp(ikr)/(ikr)$ can be neglected. The resulting $\sin(kr)/(kr)$ interaction between the atoms leads to an analytical solution from which the different observables can be computed. In particular, the intensity in a direction of angle $\theta$ reads (see Ref.~\cite{Bachelard:2011} for details)
\begin{eqnarray}
\frac{I(\theta)}{I_{10}}&=&
\sum_{n=0}^\infty \frac{(2n+1)\lambda_n}{4\delta^2+(1+\lambda_n)^2}\nonumber
\\&&+\left|\sum_{n=0}^\infty \frac{(2n+1)\lambda_n}{2\delta+i(1+\lambda_n)}P_n(\cos\theta)\right|^2, \label{I:sine}
\end{eqnarray}
where $I_{10}$ is the resonant single atom intensity (for $\delta=0$). The $\lambda_n$ are scattering eigenvalues given by
\begin{equation}
\lambda_n=N\sqrt{\frac{\pi}{2(kR)^2}}e^{-(kR)^2}I_{n+1/2}((kR)^2)
\label{lambda:n}
\end{equation}
with $I_n(x)$ is the $n$th modified Bessel function. Thus the scattered intensity is the sum of an incoherent isotropic contribution (first term of the r.h.s. of Eq.~\eqref{I:sine}), proportional to $N$, and a coherent contribution proportional to $N^2$, directed mainly in the forward direction.


{\it Shadow effect.---}
Let us first consider the isotropic (background) part of the intensity
\begin{equation}
\frac{I_\bg}{I_{10}}=\sum_{n=0}^\infty \frac{(2n+1)\lambda_n}{4\delta^2+(1+\lambda_n)^2}.
\label{I:iso}
\end{equation}

For large clouds ($kR\gg1$), the modes which contribute significantly to the scattering correspond to $n<kR$~\cite{Hulst:1981}, and their eigenvalue can be approximated by
\begin{equation}
\lambda_n\approx\frac{b_0}{4}\exp\left[-\frac{(n+1/2)^2}{2(kR)^2}\right].
\label{lambda:gauss}
\end{equation}
It is then convenient to treat the spectrum of the cloud as a continuum by defining $\eta=n+1/2$ and using the substitution $$\sum_{n=0}^\infty(2n+1)\rightarrow 2\int_0^\infty \eta d\eta,$$ along with the eigenvalues $\lambda(\eta)=(b_0/4)\exp[-\eta^2/2(kR)^2]$. This leads to
\begin{eqnarray}
\frac{I_\bg}{I_{10}}&=&2\int_0^\infty \frac{\eta\lambda(\eta)}{4\delta^2+[1+\lambda(\eta)]^2}d\eta \nonumber
\\ &=& N\frac{2}{\delta b_0}\arctan\left[\frac{\delta b_0/2}{1+4\delta^2+b_0/4}\right].
\label{I:iso:cont}
\end{eqnarray}
This expression can be rewritten as
\begin{equation}
\frac{I_\bg}{NI_{1}}=\frac{2}{\delta b}\arctan\left[\frac{\delta b/2}{1+b/4}\right].
\end{equation}
Then the contribution of the isotropic radiation to the RPF can be deduced:
\begin{equation}
\frac{F_\bg}{F_1}=
\frac{2}{\delta b}\arctan\left[\frac{\delta b/2}{1+b/4}\right],\label{F:iso}
\end{equation}
In the limit of small optical thickness $b\ll 1$, and assuming also $\delta b\ll 1$, one recovers
\begin{equation}
\frac{F_\bg}{F_1}\approx 1-\frac{b}{4},\label{eq:Fmodalbg}
\end{equation}
which corresponds to the first correction of the shadow effect [Eq.~\eqref{eq:Fshad}].

{\it Forward lobe.---}
The forward contribution of the intensity is given by the second term in Eq.~\eqref{I:sine}, which needs to be inserted in Eq.~\eqref{eq:IF}.
This leads to calculating the product between different modes, which is realized using the following formula:
\begin{eqnarray}
\int_0^\pi &d\theta\sin\theta(1-\cos\theta)
P_n(\cos\theta)P_m(\cos\theta)\label{iden:P}
\\ &=\dfrac{2\delta_{n,m}}{2n+1}-
2\dfrac{(m+1)\delta_{n,m+1}+m\delta_{m,n+1}}{(2m+1)(2n+1)}.\nonumber
\end{eqnarray}
In the above expression, the first r.h.t. corresponds to the force associated to the absorption of the light, whereas the second one is associated to the emission. One thus obtains an absorption force,
\begin{equation}
\frac{F_\abs}{F_1}=\frac{1+4\delta^2}{N}\sum_{n=0}^\infty
\frac{(2n+1)\lambda_n^2}{4\delta^2+(1+\lambda_n)^2},\label{force:lobe4}
\end{equation}
which, in the continuous spectrum approximation, turns into
\begin{eqnarray}
\frac{F_\abs}{F_1}&\approx& \frac{2(1+4\delta^2)}{N}\int_0^\infty
\frac{\eta\lambda^2(\eta)}{4\delta^2+(1+\lambda(\eta))^2}d\eta,\label{force:lobe6}
\\ &\approx& \frac{2}{b}
\ln\left[1+\frac{b}{2}\left(1+\frac{b_0}{8}\right)\right]
-\frac{2}{\delta b}\arctan\left[\frac{\delta b/2}{1+b/4}\right]\nonumber
\end{eqnarray}
The force associated to the emission of the light in a  forward cone of aperture $\sim 1/kR$ is provided by the last term in \eqref{iden:P} in conjunction with \eqref{I:sine}. After reorganizing the sums, one obtains
\begin{eqnarray}\nonumber
\frac{F_\emi}{F_1}&=&-\frac{1+4\delta^2}{N}\sum_{n=0}^\infty
\frac{2(n+1)\lambda_n\lambda_{n+1}}{4\delta^2+(1+\lambda_n)^2}
\\ && \times \left[
\frac{4\delta^2+(1+\lambda_n)(1+\lambda_{n+1})}{4\delta^2+(1+\lambda_{n+1})^2}\right]. \label{force:lobe7}
\end{eqnarray}
Using the low optical thickness limit and Eq.~\eqref{lambda:gauss}, the dependence on the detuning disappears and the emission force writes
\begin{eqnarray}
\frac{F_\emi}{F_1}&\approx&-\frac{2}{N}\sum_{n=0}^\infty (n+1)\lambda_n\lambda_{n+1}
\\ &\approx&-\frac{b_0}{8}\exp[-1/2(kR)^2]\label{force:gauss}
\end{eqnarray}
Finally, the sum of the isotropic background contribution \eqref{eq:Fmodalbg} and of the absorption \eqref{force:lobe6} and emission \eqref{force:gauss} terms, using the $kR\gg1$ and $b\ll1$ limits, leads to the net RPF:
\begin{equation}
\frac{F_\mathrm{modal}}{F_1}\approx
1-\frac{b}{4}+\frac{b_0}{16(kR)^2}.
\label{force:lobe:linear}
\end{equation}


\end{document}